# Study of Electric Quadrupole Perturbation at Multiple Probe Sites in Hf-doped Rutile in a Single Perturbed γ-γ Angular Correlation Measurement

D. Banerjee[1*], S. K. Das[1] and S. V. Thakare[2]

[1]*Radiochemistry Division (BARC), VECC, 1/AF, Bidhan Nagar, Kolkata-700064, India.*
[2]*Radiopharmaceuticals Division, BARC, Mumbai-400085, India.*

**Abstract:**

Two different probes, viz., $^{111}$In/$^{111}$Cd and $^{181}$Hf/$^{181}$Ta, simultaneously have been incorporated in both pure rutile and Hafnium (Hf) -doped rutile Titanium Dioxide (TiO$_2$) systems following a soft chemical method. The local information around two different probe sites in the same sample could be extracted in a single Time Differential Perturbed Angular Correlation (TDPAC) measurement using LIST mode data acquisition. The two probes were incorporated in the rutile samples by co-precipitation technique and then annealed at 1273K for 10h to get the rutile structure. The rutile structure was identified by the hyperfine parameters, viz., quadrupole frequency ($\omega_Q$) and asymmetry ($\eta$). At $^{111}$Cd site, $\omega_Q$=16.8(1) Mrad/s and $\eta$=0.19(1) while at $^{181}$Ta site, $\omega_Q$=127.2(2) Mrad/s and $\eta$=0.56(1). Due to the doping of Hf at 5 and 10 atom%, the sample partially lost the purity of the crystal and the fact has been reflected at both the probe sites.

*Corresponding author: Email: dbanerjee@vecc.gov.in*
*Tel: +91 33 23183280; Fax: +91 33 23346871*

## 1. Introduction

In the field of TDPAC [1-3], there have been a number of probes used to study different chemical systems. The probes have been chosen based on the nature of the system to be studied. Till date, a single TDPAC measurement has involved a single probe only. However, a single system could have been studied with multiple probes by carrying out separate TDPAC measurement for each probe. In this context, it is quite convenient to use multiple probes for multiple sites available in the system to be studied in a single TDPAC experiment. It is particularly important while studying systems with multiple elements, viz., intermetallic alloys, perovskites etc. This technique could avoid any complexity which could have arisen during sample preparation or in other words, it would assure the equivalent experimental condition. None the less, it saves the acquisition time too. The prerequisite to use multiple probes in a single experiment lies in the fact that the relevant gamma energy lines of the probes are well separated by the detectors and the lifetimes of the intermediate level for the different probes do not differ to a large extent.

$TiO_2$ has found several applications in the field of photocatalysis and photochemistry [4-6]. It is a wide band gap (=3.2 eV for anatase and 3.0 eV for rutile phase) semiconductor [7] and these have absorption in the UV region ($\lambda \leq 380$ nm). But the solar radiation that reaches the earth's surface lies in the range of 1.0-1.9 eV. Again, it would be desirable to decrease the photocatalytic efficiency of $TiO_2$ to be used as a paint opacifier since it causes the radical-formation resulting in the degradation of the organic binders in paint. Hence, the optical properties of $TiO_2$ are to be tuned properly for its use in different applications. Again, for another important application of the photocatalytic property of $TiO_2$ towards the hydrogen production, it is required to adjust band-gap of $TiO_2$. Doping as an effective way to modify the band gap. A recent review article [8] provides information about the methods of preparation of doped-$TiO_2$ with metallic and nonmetallic species, including various types of dopants and doping methods. Again the effect of doping on the optical and photocatalytic properties of $TiO_2$ has also been furnished [9] with possible explanation of those effects. For this purpose, $TiO_2$ is doped with transition metal ions (Fe, Mn, Ni, Cr etc.) and non-metals (N, S, C, B etc.). Recently, $TiO_2$ doped with multiple dopants has also attracted attention and the dopant-dopant/dopant-host interaction could be an important study to modify the optical properties of $TiO_2$-based photocatalysts [10]. Such study of $TiO_2$ systems doped with multiple dopants could be done very effectively with microscopic tool like TDPAC. In those studies, it could be pretty effective to use multiple probes as well in order to probe multiple dopant sites and it is always preferable if it could be done in a single experiment and with a single sample. The present work is an effort to demonstrate such use of multiple probes in a single TDPAC experiment.

Two different probes, viz., $^{111}$In/$^{111}$Cd and $^{181}$Hf/$^{181}$Ta, have simultaneously been used in the experiment. Both the probes were incorporated in pure and 5% Hf-doped $TiO_2$ samples and the local structure at the two probe sites has been extracted. The known TDPAC parameters for the two probes in rutile structure [11] have been used to identify the different oxide phases in the present study. The $^{181}$Hf/$^{181}$Ta and $^{111}$In/$^{111}$Cd probes are added at a very dilute concentration (mole fraction ~ $10^{-11}$) and expected to occupy the Ti site in the rutile structure [11]. In case of Hf-doped $TiO_2$, Hf is present at 5 and 10 atom% level in the present case. Considering the ionic radii and most stable oxidation states of Hf and Ti atoms, it is expected that Hf should replace Ti site in the rutile structure. But, $TiO_2$ and $HfO_2$ have quite different crystal structures, viz., tetragonal and monoclinic respectively. Ti atom is coordinated to six oxygen atoms in $TiO_2$ while Hf atom is bonded to seven oxygen atoms in $HfO_2$. Hence, the solubility of Hf in rutile structure might have got an upper limit and a separate phase formation is quite possible beyond a certain atom% of Hf. In order to identify the second phase, present in a minuscule amount, the TDPAC having atomic scale resolution could be an effective tool. Again it might be the case that the probe used in the study could not get incorporated into the second phase. In such cases also, the use of multiple probes in the same sample and in the same experiment could minimize the probability of loosing such minute details of the sample.

## 2. Experiment

*2.1. Sample Preparation*

Titanium (IV) isopropoxide (Aldrich, 99.999%) was used as the precursor for $TiO_2$ and Hafnium (IV) chloride (Fluka, ≥97%) was used for Hf-doping. $^{181}$Hf/$^{181}$Ta probe was obtained by neutron-irradiation of Hf-oxychloride at DHRUVA reactor of Bhabha Atomic Research Centre, Mumbai, India. A stock solution of this activity was prepared by dissolving irradiated Hf-oxychloride in dilute acid and used for subsequent sample preparation. $^{111}$In/$^{111}$Cd probe was produced from the natural silver (Ag) foil using $^{nat}$Ag (α, xn) reaction. The reaction was performed using 30 MeV α-beam delivered by the AVF Cyclotron of Variable Energy Cyclotron

Centre (VECC), Kolkata, India at a current of 1μA. The $^{111}$In activity was then radiochemically separated from the Ag-matrix [12] and this carrier-free $^{111}$In-activity was used for further TDPAC measurements. Rutile TiO$_2$, doped with 5 and 10 atom% of Hf, was prepared from the Ti-isopropoxide precursor by co-precipitation method [13] followed by annealing at 1273K for 10h. The PAC probes were incorporated during the preparation of the rutile system.

*2.2. TDPAC Measurement*

In case of TDPAC, it requires a γ-γ cascade fed by the decay of a parent isotope. The parent isotope is produced by a nuclear reaction and then decays by particle or γ-ray emission to produce the daughter isotope. Then the daughter atom, inside the matrix under study, acts as an "wound-up spy" [14] to transfer information of its host matrix while returning to its ground state. The intermediate state of the cascade should have a lifetime in the range of 10-1000 ns. This is a crucial parameter as the intermediate state must have a sufficiently long lifetime to feel the perturbation. Again the extranuclear field must be significant enough to interact with the quadrupole moment of the intermediate state. The interaction of the nuclear electric quadrupole moment (*Q*) with an electric field gradient (EFG) leads to a perturbation or attenuation in the angular correlation pattern.

TDPAC technique is based on the hyperfine interaction technique and relies on the perturbation of the angular correlation of the cascade gamma rays emitted by the probe nucleus. TDPAC measurements were carried out using 133-482keV γ-γ cascade of $^{181}$Ta after the beta (β) decay of $^{181}$Hf parent and 171-245keV γ-γ cascade of $^{111}$Cd after the electron capture (EC) decay if $^{111}$In parent. The TDPAC spectrometer consists of three LaBr$_3$(Ce) detectors coupled to fast XP2020URQ photomultiplier tubes and data were collected in list mode using CAMAC data acquisition system [11].

In TDPAC technique, the intermediate state is split into different non-degenerate m-states with eigenvalues proportional to the quadrupole frequency $\omega_Q$ which is given by:

$$\omega_Q = \frac{eQV_{zz}}{4I(2I-1)\hbar}$$

Where, *I* is the spin of the intermediate level. Due to the above interaction, the angular correlation function is attenuated by a factor called as the perturbation factor G$_2$(t).

From the coincidence measurement of the 133-482 keV and 171-245 keV cascades at 90° and 180°, the obtained perturbation factor ($G_2(t)$) was fitted to the following equation:

$$G_2(t) = a_0 + \sum_{n=1}^{3} a_n \exp(-\omega_n \delta t) \times \exp(-1/2\, \omega_n^2 \tau^2) \cos(\omega_n t)$$

The exponential damping terms attribute to the finite resolving time characterized by a Gaussian distribution with standard deviation $\tau$ and the Lorentzian frequency distribution with relative width parameter $\delta$. The coefficients $a_n$ have dependence on the nuclear radiation parameters and the asymmetry parameter $\eta = (V_{xx} - V_{yy})/V_{zz}$.

The least square fitting was carried out using the code WINFIT [15] to obtain the TDPAC parameters: $\omega_Q$, $\eta$ and $\delta$. When there is more than one well defined lattice site that the probe atom could occupy, the perturbed angular correlation function is to be adjusted for each site. Then the total perturbation function with different site fractions $f_i$ is given by the linear superposition of $G_2(t)$ for each probe environment:

$G_2(t) = \sum_{i=1}^{n} f_i G_2^i(t)$ with the normalization factor $\sum_{i=1}^{n} f_i = 1$.

## 3. Results and Discussion

In order to emphasize the excellent energy resolution of the LaBr$_3$(Ce) detectors, the energy spectrum for the sample containing both the probes is shown in Fig. 1. It is seen that all the energy peaks are well separated showing the excellent energy resolution of the LaBr$_3$(Ce) detector. The energy gates could be given in the respective γ-peaks to obtain the gated time spectra shown in Fig. 2 during post-acquisition analysis. The upper one is the energy gate given in the 171-245 keV cascade of $^{111}$Cd after the EC decay of $^{111}$In and lower one is the 133-482 keV cascade of $^{181}$Ta after the β-decay of $^{181}$Hf. As the relevant gamma lines are well separated, the gates could easily be given and this is a prerequisite to use multiple probes in a single TDPAC experiment. The TDPAC spectra for pure rutile TiO$_2$ and 10% Hf-doped rutile with $^{111}$Cd probe are shown in Fig. 3. The A$_2$G$_2$(t) spectra are shown in the left side of the figure and the corresponding cosine transform is shown in the right side. The corresponding TDPAC spectra for 5% Hf-doped rutile with $^{111}$Cd and $^{181}$Ta probes are shown in Fig. 4. The TDPAC parameters for different rutile samples at $^{111}$In/$^{111}$Cd probe site are given in Table 1 and those at $^{181}$Hf/$^{181}$Ta probe site are given in Table 2. In Fig. 4, it is to be noted that the time range in axis is not same for the two probes and chosen on the basis of the lifetime of the intermediate level.

The modulations over the exponential decay in two time spectra, shown in Fig. 2, appear to be different in two energy gates. Such time spectra for multiple energy gates could be extracted with a prerequisite that the energy peaks are well separated by the detector used in the study. By this LIST mode method using CAMAC data acquisition technique, multiple probes could be introduced in the same sample and the information at each probe site could be extracted in a single TDPAC experiment. It would have several advantages over the conventional single probe technique, viz., it would maintain identical experimental condition, save a lot of time by not carrying out TDPAC experiment for each probe separately, help in even cross-checking the results and specially, deliver a useful information while studying a system with multiple types of atomic sites.

Rutile $TiO_2$ has a tetragonal structure with space group $P4_2/mnm$. In this, the Ti atom is surrounded by six O atoms as the nearest neighbors in an octahedral geometry. In the next layer there are eight Ti atoms in the corner of the tetragon. In case of pure rutile compound, the fitting could be done with a single set of TDPAC parameters, as shown in the upper panel of Fig. 3 for $^{111}$In/$^{111}$Cd probe. The frequency values for pure rutile were found to be 127.2(2) and 16.8(1) Mrad/s in case of $^{181}$Ta and $^{111}$Cd probes respectively. However, the data for doped rutile could be fitted with more than one component, as depicted in Table 1 and 2, for the two probes. One component was obviously attributed to the pure rutile site with above frequency values. The other site with large frequency distribution was attributed to an amorphous phase. The population of this amorphous phase increased with increasing Hf-content. The width of the frequency distribution for the rutile structure at $^{111}$Cd-site also increased from 4.3(4)% to 6.6(1.0)% as the Hf-content was changed from 5% to 10%. The effect of Hf-doping on the rutile structure appeared to be quite different from the effect of Zr-doping in the rutile structure as observed using $^{181}$Hf/$^{181}$Ta probe [13]. It is to be noted that the inherent crystallinity of the pure rutile sample at the two probe sites is not identical. In case of $^{111}$Ta site, the frequency distribution is 0.6(1)% while it is 1.2(1)% in case of $^{111}$Cd-site. Again, on Hf-doping at 5 atom% level, the distribution increases from 1.2(1) to 4.3(4)% at $^{111}$Cd site but the similar increase at $^{181}$Ta site is from 0.6(1) to 1.1(5), as seen in Table 1 and 2. This suggests that there exists a metal-metal (M-M) interaction between the central atom (i.e., probe atoms: $^{181}$Ta or $^{111}$Cd) and the metal atom in the next nearest neighbor. It also suggests that the M-M interaction between probe atom and host atom (here Ti) is different from that between probe atom and dopant (here Hf). The effect of Hf-doping on the frequency distribution at the two probe sites indicates that the Cd-Hf interaction is more pronounced in the next nearest neighbor than the Ta-Hf interaction.

From the Table 2, it is seen that the TDPAC data for 5%Hf-doped rutile at $^{181}$Ta-site could be fitted with three different set of parameters. One set corresponds to pure rutile site, second set with high δ-value corresponds to an amorphous phase and the third one corresponds to a new set of data. The amorphous phase has the frequency value of ~212 Mrad/s which could be correlated to the similar site observed at $^{111}$Cd site after correcting for the $Q$-value and the effective charge. The third site with frequency~100Mrad/s and η~0.75 could not be attributed to pure monoclinic HfO$_2$ site which has a frequency value of 124.6 Mrad/s and η~0.36 [16]. So, this site was presumably attributed to the monoclinic HfO$_2$ site with one point defect. This conjecture was made on the basis of the fact that the removal of one O-atom from the seven O-atoms around the Hf-atom in HfO$_2$ structure would decrease the effective charge and increase the asymmetry of the site resulting in a decrease of quadrupole frequency and increase in η-value respectively.

A simple numerical calculation indicates that one Ti atom is surrounded by 8 Ti atoms. To replace at least one Ti by a Hf atom required 1:8 ration of Hf:Ti atom i.e. ~12 %Hf. So 5% doping may cause more than 75% lattice remains pure rutile and rest is having one impurity atom in the next neighbor. M-M interaction may not thus fully explain the observation of only 33% pure rutile structure. Alternatively the observation can be explained in terms of partial amorphisation of rutile phase by doping with Hf. It is interesting to see that only 5% doping can amorphise almost 67% of the lattice leaving only 33% pure rutile phase. However, 10% doping enhances the amorphisation to 80%. This marginal enhancement of the amorphous phase may be due to the formation of HfO$_2$ phase (with defect) at a larger doping of Hf. This has been observed in the case of $^{181}$Hf probe as mentioned in the Table 2. Non-observation of such defect HfO$_2$-site with $^{111}$In probe is due to the fact that In could not go into the HfO$_2$ phase, for which the reason is not clearly understood. So the above study demonstrates the utility of using more than one probe in a single TDPAC experiment. The information obtained at the two probe sites in the present work is not identical and rather complimentary to each other.

## 4. Conclusion

The present study demonstrates the use of dual probe in a single TDPAC experiment and highlights the advantage of this technique over the conventional single probe TDPAC experiment. The two probes, viz., $^{181}$Hf/$^{181}$Ta and $^{111}$In/$^{111}$Cd, were used in the present study and the information obtained at the two probe sites has been furnished while studying pure and doped rutile $TiO_2$. The effect of Hf-doping on attainment of rutile structure under a particular experimental condition has been explored with the above two probes simultaneously in a single TDPAC experiment. The data were collected in list mode and information at each probe site was separately extracted in the post acquisition period. It has been observed that the doping of rutile with Hf introduced a separate amorphous phase at the cost of pure rutile structure and the percentage of this amorphous phase increased with increasing Hf-doping. Both the probes were found to be in pure rutile and amorphous phase. The $^{181}$Hf/$^{181}$Ta probe was additionally found to be in a monoclinic $HfO_2$ phase with point defect. This extra information obtained at the $^{181}$Ta site was missing at $^{111}$Cd site. Thus the present study successfully describes the dual probe TDPAC measurement and its advantages over the single probe technique.

**Acknowledgement**

The authors are sincerely thankful to the cyclotron group members, VECC, Kolkata . The authors also heartily thank Dr. P. K. Pujari, Head, Radiochemistry Division, BARC and Dr. R. Acharya, Head, Accelerator Chemistry Section for their keen interest in this work.

**Table 1:** TDPAC parameters for rutile samples with [111]In/[111]Cd probe

| Rutile samples | $\omega_Q$(Mrad/s) | $\eta$ | $\delta$(%) | Population(%) | Remarks |
|---|---|---|---|---|---|
| Pure | 16.8 (1) | 0.19(1) | 1.2(1) | 100 | Rutile site |
| 5%Hf doped | 16.83(8) | 0.22(1) | 4.3(4) | 33 | Rutile site |
|  | 25.2(2.6) | 0.67(14) | 33.2(5.2) | 67 | Amorphous |
| 10%Hf doped | 17.2(2) | 0.24(3) | 6.6(1.0) | 20 | Rutile site |
|  | 24.1(2.7) | 0.86(28) | 28.7(3.9) | 80 | Amorphous |

**Table 2:** TDPAC parameters for rutile samples with [181]Hf/[181]Ta probe

| Rutile samples | $\omega_Q$(Mrad/s) | $\eta$ | $\delta$(%) | Population(%) | Remarks |
|---|---|---|---|---|---|
| Pure | 127.2 (2) | 0.56(1) | 0.6(1) | 100 | Rutile site |
| 5%Hf doped | 126.2(2) | 0.57(1) | 0.8(4) | 27 | Rutile site |
|  | 224.9(1.5) | 0.34(1) | 11.2(5) | 56 | Amorphous |
|  | 103.7(4) | 0.75(1) | 3.5(5) | 17 | Defect site |

**Figures:**

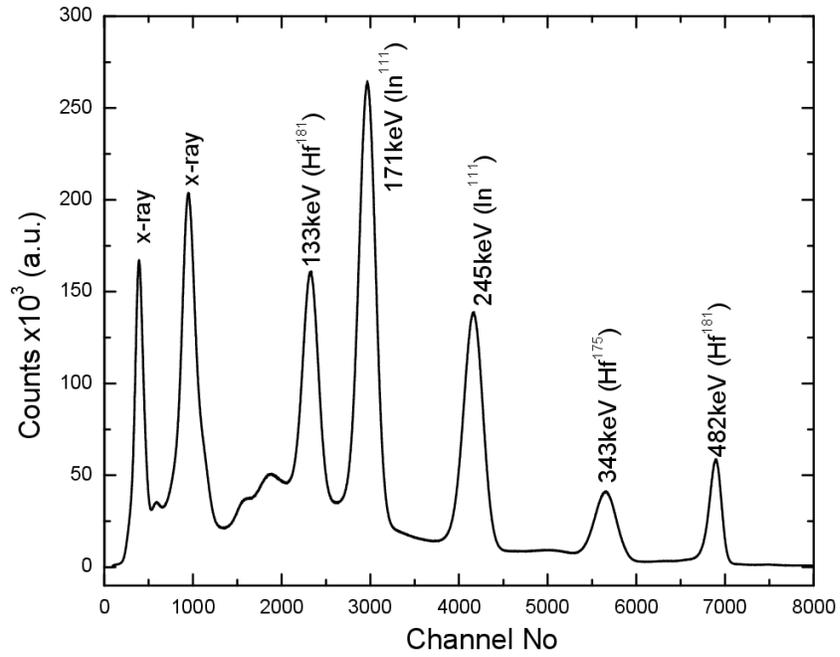

**Fig. 1:** Energy spectrum of LaBr$_3$(Ce) detector for ($^{111}$Cd+$^{181}$Ta) probes

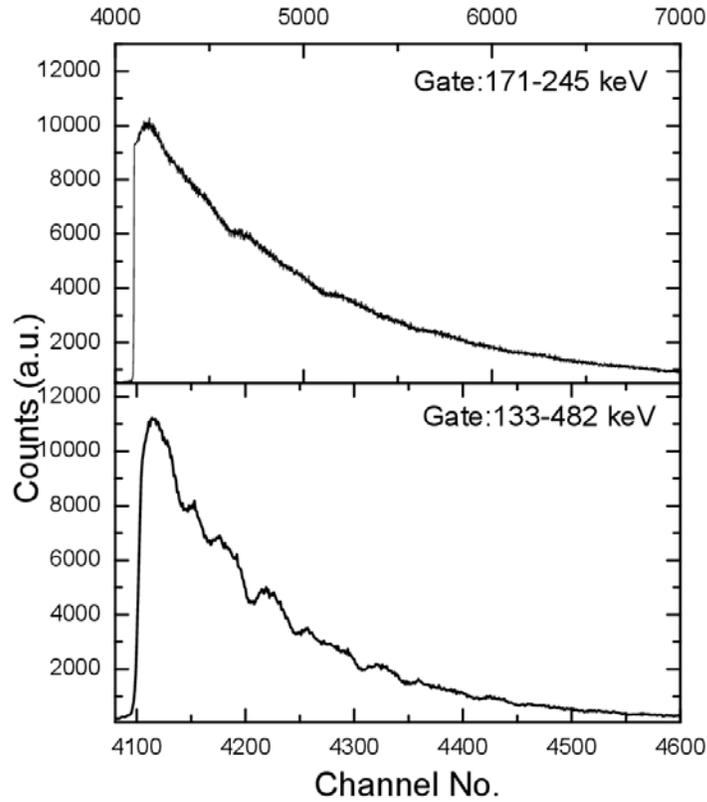

**Fig. 2:** Two gated time spectra extracted from the same time spectrum for the two probes

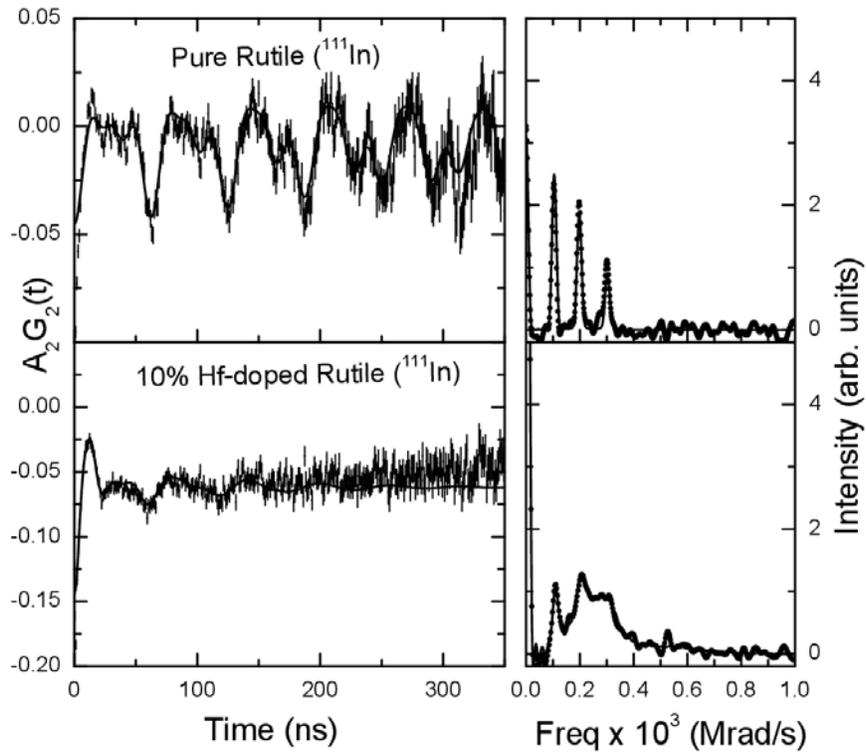

**Fig. 3:** TDPAC spectra for pure and 10% Hf-doped rutile samples with $^{111}$Cd probe

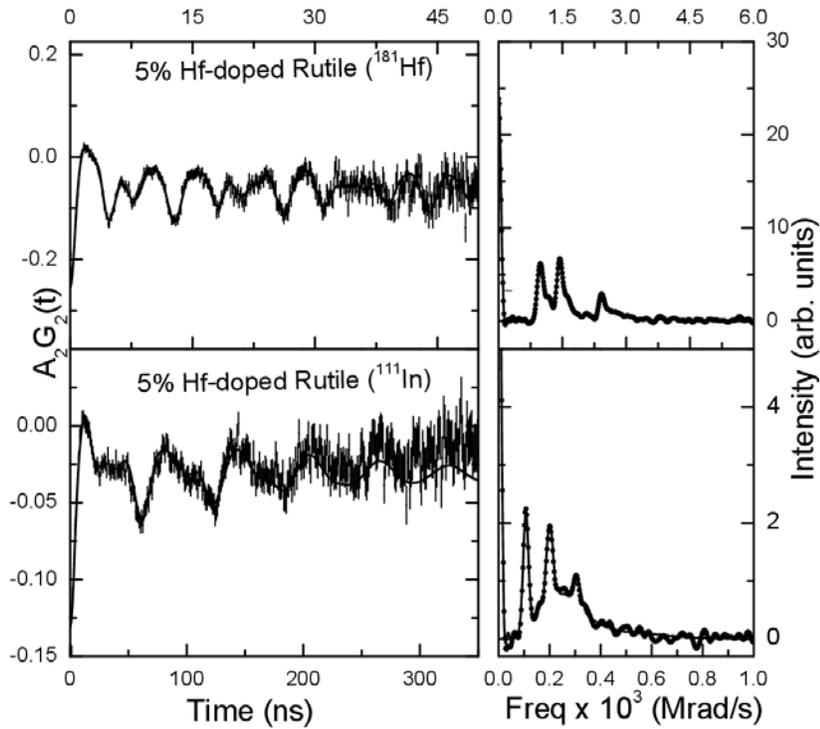

**Fig. 4:** TDPAC spectra for 5% Hf-doped rutile samples with two probes